\documentclass[pdflatex,sn-mathphys,iicol]{sn-jnl}

\jyear{2021}%

\theoremstyle{thmstyleone}%
\theoremstyle{thmstyletwo}%

\theoremstyle{thmstylethree}%

\begin{document}

\title[Article Title]{The crucial role of elasticity in regulating liquid-liquid phase separation in cells}

\author*[1,2]{\fnm{Mrityunjay} \sur{Kothari}}\email{mjayk@mit.edu}

\author*[1,3]{\fnm{Tal} \sur{Cohen}}\email{talco@mit.edu}

\affil[1]{\orgdiv{Department of Civil and Environmental Engineering}, \orgname{Massachusetts Institute of Technology}, \orgaddress{\street{77 Massachusetts Ave}, \city{Cambridge}, \postcode{02139}, \state{MA}, \country{USA}}}

\affil[2]{\orgdiv{Department of Mechanical Engineering}, \orgname{University of New Hampshire}, \orgaddress{\city{Durham}, \postcode{03824}, \state{NH}, \country{USA}}}

\affil[3]{\orgdiv{Department of Mechanical Engineering}, \orgname{Massachusetts Institute of Technology}, \orgaddress{\street{77 Massachusetts Ave}, \city{Cambridge}, \postcode{02139}, \state{MA}, \country{USA}}}


\abstract{{Liquid-liquid phase separation has emerged as a fundamental mechanism underlying intracellular organization, with evidence for it being reported in numerous different systems.
However, there is a growing concern regarding the lack of quantitative rigor in the techniques employed to study phase separation, and their ability to account for the complex nature of the cellular milieu, which affects key experimentally observable measures, such as the shape, size and transport dynamics of  liquid droplets.
Here we bridge this gap by combining recent experimental data with theoretical predictions that capture the subtleties of nonlinear elasticity and fluid transport. We show that within a biologically accessible range of material parameters, phase separation is highly sensitive to elastic properties and can thus be used as a \textit{mechanical switch} to rapidly transition between different states in cellular systems.
Furthermore, we show that this active mechanically mediated mechanism can drive transport across cells at biologically relevant timescales and could play a crucial role in promoting spatial localization of condensates; whether cells exploit such  mechanisms for transport of their constituents, remains an open question.}}

\keywords{Liquid-Liquid Phase Separation, Ostwald ripening, Elasticity, Front Propagation}



\maketitle

\section{Introduction}\label{sec1}

Liquid-liquid phase separation (LLPS) has emerged as a fundamental mechanism by which eukaryotic cells organize themselves into membraneless compartments {called biomolecular condensates} that carry out important cellular functions \citep{brangwynne2009germline, hyman2014liquid, alberti2019considerations}. Despite its recognition over the last decade as a means of self-organization in biology, there is an emerging skepticism regarding the supporting evidence for LLPS \citep{mcswiggen2019evaluating, Leslie336}.
A major criticism levelled at the current approaches to study LLPS is their reliance on qualitative measures and the lack of concomitant quantitative modeling.
For instance, Fluorescence Recovery After Photobleaching (FRAP) -- a technique used commonly to ascertain the liquid-like nature of {biomolecular condensates} -- measures the time it takes for photo bleached molecules to diffuse out of these {liquid droplets}. 
However, the observed diffusion times from FRAP studies \citep{mcswiggen2019evaluating} show a huge spread (from a few seconds to minutes) and are likely to be sensitive to the {droplet's complex environment (such as the elasticity of biopolymer networks in the medium surrounding the droplet)}, yet this connection is poorly understood.
Further, qualitative approaches are not infallible to experimental limitations (like diffraction limit), post-processing induced artifacts, and presence of alternate mechanisms that lead to similar outcomes. 
Given the small size of {biomolecular condensates} and the complex nature of the physiological environment inside cells, purely qualitative observations can be misleading without accounting for the various physical effects that are significant at cellular length scales.
Therefore, there is a need to supplement the current methods of investigation with rigorous quantitative analyses that establish the crucial link between {the medium's} material properties and the outcome of LLPS, and can help guide and interpret future studies.
    
Recent experiments have revealed that the sea of \textit{elastic} network contained in cells -- which impart them an elastic-solid like mechanical resistance to deformation -- play a very important role in LLPS by providing elastic resistance to the growth of {biomolecular condensates} \citep{zhang2020mechanical, lee2021chromatin, boddeker2021non, fernandez2021putting}.
This mechanism may hold answers to why LLPS inside the cellular milieu leads to tightly regulated {condensate} sizes and associated dynamics, unlike the classic example of oil-water phase separation.

{Synthetic polymer based systems have provided a simpler \textit{droplet-in-polymer} analog to \textit{condensate-in-cell} systems to experimentally and theoretically study the effects of elastic resistance \citep{style2018liquid, rosowski2020elastic, rosowski2020elastic2, ronceray2021liquid, vidal2021cavitation, mukherjee2021statistical, biswas2021thermodynamics,paulin2021fluid, wei2020modeling}.
These systems have raised some intriguing questions.}
Existing studies frequently treat droplet growth as a cavitation process \citep{ball1982discontinuous}, which assumes a constant, {size-independent pressure inside the droplet}, and thus does not energetically distinguish between a single large droplet and multiple small droplets (overall droplet volume being the same). Then why do we see the latter and not the former?
Furthermore, how does elasticity affect the commonly measured experimental quantities, such as size of the droplets, their coarsening dynamics, and their spatial localization?

{In our previous work on liquid-liquid phase separation inside synthetic polymers \citep{kothari2020effect}, we showed that the elastic resistance imposed by the medium delays phase separation, arrests droplet growth, and can even drive flux of liquid from regions of high stiffness to low stiffness.
For the specific problem at hand, our analysis in \citep{kothari2020effect} was restricted to the assumption of infinite domains and employed a simple transport law where liquid flux was assumed to be driven by concentration gradients.}

{To quantitatively understand the role of elasticity in intracellular phase separation, where the material properties and length scales differ substantially from synthetic systems studied so far in the literature, there is a need to port the knowledge from synthetic systems to biological systems and to address the above limitations.}
In particular, in recent years there has been an accumulation of evidence that strain stiffening plays an important role  in various functions of the cell \cite{li2021nonlinear,han2018cell,van2016strain,hu2019high}. {Could this effect be crucial, even at the cellular length scales,} in explaining the size regulation of membraneless organelles?
Motivated by these questions, we develop a theoretical model that builds on \citep{kothari2020effect}, and apply it to explain experimental observations in relevant cellular systems. We then extend the analysis beyond the experimental range to broaden our understanding of the constitutive sensitivities.

\section{Model}\label{sec2}

{The model presented in this section closely follows the authors' previous work \citep{kothari2020effect} with a key point of departure: in this work we employ a generalized kinetic model by prescribing the flux as proportional to its thermodynamic conjugate, i.e. gradients in chemical potential\footnote{{The chemical potential is defined here as the change in Gibbs free energy of a system when a molecule is removed or added to it at constant temperature and pressure.}}, as opposed to prescribing the flux proportional to concentration gradients. 
This kinetic law allows us to model a much broader range of material heterogeneities and is capable of describing both \textit{downhill} (from higher to lower concentration) and \textit{uphill} diffusion (from lower to higher concentration), thereby widening the applicability of the theory to real-world systems.}

Our model system {comprises} an elastic network of cross-linked polymer, permeated by a liquid mixture that is made of the following two components: free liquid of the corresponding un-crosslinked or partially crosslinked polymer chains \citep{jensen2015wetting} (denoted by A), and another liquid of a different species (denoted by B).
The model system is a {simplified representation} of cellular milieus that show phase separation. 
{For example, in cells,} proteins and RNAs are mixed inside the cytoplasm, which contains elastic network as well as many other liquids. 
Proteins and RNAs phase separate to create P granules, that segregate but coexist with cytoplasm \citep{brangwynne2009germline, hyman2014liquid}. 
Similarly, inside the nucleus, the nucleolar protein fibrillarin (FIB-1) phase separates from the nucleoplasm to create nucleoli \citep{berry2015rna}.

Typically, there are two main time scales associated with phase separation in heterogeneous materials.
{\textit{Short timescale:}} as the mixture {(supersaturated with liquid B) is stimulated, droplets of liquid B} nucleate and grow inside the elastic matrix before achieving a quasi-equilibrium size, which is governed by a balance between the local chemical and elastic properties of the mixture.
If the nucleation and growth process is rapid, the global heterogeneity does not influence the local behavior of the droplets, which involves only short range migration of liquid and thus only depends on local elastic properties. 
{\textit{Long timescale:}} heterogeneity can dominate transport and drive long-range migration of liquid across the material at a longer timescale.
In the current work, we will restrict our attention to scenarios where the timescales are well separated. This allows us to treat nucleation and growth separately from the ensuing dynamics.
 
We make the following assumptions in our model.
We consider situations in which the matrix does not swell appreciably in the entire process, and thus changes in elastic energy due to the swelling of the matrix can be neglected. 
Accordingly, we model the process as mixing between liquids A and B, where the cross-linked network only provides elastic resistance to droplet growth, and only liquid B can migrate spatially.
Restricting our attention to the dilute limit, we also ignore the strain energy due to elastic interaction among droplets,
{and we assume the droplet distribution to be monosdisperse.}

The concentration of liquid B in the mixture is denoted by $\phi$, and is defined with respect to the matrix volume (i.e. excluding the droplet volume).  
Concentration of the liquid in the droplet phase is denoted by $\phi_D =\frac{4\pi}{3}r^3 n_d$, where $r$ is the radius of the droplets, and $n_{d}$ is the number of droplets per unit volume; $\phi_D$ is defined with respect to the total volume within a representative element.
 
As a supersaturated mixture of initial concentration $\phi^{sup}$ phase-separates, the droplets nucleate and start to grow in the elastic matrix.
Phase separation is driven by the lowering of the mixing free energy ($\Delta \bar{G}_{mix}$). At later stages of phase separation, reduction of surface energy ($\Delta \bar{G}_{sur}$) drives coarsening; however, the growth and coarsening of droplets incurs substantial elastic energy ($\Delta \bar{G}_{el}$). 
This energy competition drives the system to choose a droplet size that minimizes the total free energy. In terms of driving forces, at equilibrium, the chemical potential of liquid B $(\mu)$ in the matrix balances the chemical potential of the droplet $(\mu_D)$.

The change in total free energy density of the system, calculated per unit  volume, can be expressed as the sum  of contributions from mixing, surface, and elastic energies
\begin{multline}\label{eq:total-free-energy}
\Delta \bar{G}(\phi,\phi_D) =  \Delta \bar{G}_{mix}( \phi, \phi_D) + \Delta \bar{G}_{sur}(\phi_D) \\ +\Delta  \bar{G}_{el}(\phi_D), 
   \end{multline}
respectively. In this work we will use the Flory–Huggins solution theory \citep{flory1953principles} to represent changes in mixing free energy,\footnote{We assume that the number of lattice sites occupied by a single chain of the liquid $N_A \gg 1$ for simplicity.} $$ \Delta \bar{G}_{mix}= \frac{k T}{\nu_m}(1-\phi_D)(\phi \ln\phi+ \chi(T) \phi(1-\phi)),$$ where $k$ denotes the Boltzmann constant, $T$ denotes the temperature, $\nu_m$ denotes the molecular volume of liquid B, and $\chi$ denotes the Flory-Huggins parameter{, which can be inferred  by energy minimization for a given  saturation concentration of liquid B} (see Appendix A). Changes in surface energy are written as $ \Delta \bar{G}_{sur}(\phi_D)= 4\pi r^2\Gamma n_d$, where $\Gamma$ is the surface energy between the two liquids, and $\Delta \bar{G}_{el}(\phi_D)=\frac{4\pi}{3}r^3W(r) n_d$, where $W(r)$ denotes the elastic strain energy density due to growth of a single droplet (see Appendix B).

In the short timescale, before long-range migration occurs, the local concentration of the absorbed liquid, must be conserved, namely
\begin{equation}\label{eq:mass-conservation} 
    \phi^{sup} = \phi(1-\phi_D) +\phi_D,
\end{equation}
{where $\phi^{sup}$ denotes the concentration of the supersaturated mixture prior to phase separation.}
With the specific response functions chosen, the short-timescale equilibrium droplet size can now be calculated by minimizing the total free energy \eqref{eq:total-free-energy} subject to the mass conservation constraint \eqref{eq:mass-conservation}. 
Minimizing the free energy is equivalent to finding the concentration, $\phi$, that equilibriates the chemical potential of the liquid in the matrix {($\mu$)}  to the chemical potential of the liquid in the {droplet} {($\mu_D$).}  
{When the droplets and matrix are in equilibrium (minimum energy state), exchange of a molecule will not alter the total energy. In other words, their chemical potentials are the same. } The derivative 
${\rm d}\Delta \bar{G}/{\rm d} \phi =0$ gives
\begin{eqnarray}\label{eq:quasi-eqm1} 
\begin{split}
     \underbrace{kT(\ln\phi + (1-\phi) +\chi(1-\phi)^2)}_{\mu}= \quad\quad \\ \underbrace{\nu_m\left(\frac{2\Gamma}{r}+ W(r) +\frac{r}{3}W'(r) \right)}_{\mu_D},
 \end{split}
 \end{eqnarray}
where we have used ${\rm d}(\phi(1-\phi_D)+\phi_D)=0$ from \eqref{eq:mass-conservation} (see Appendix C for details).
{Here, $\mu$ {is} due to changes in free energy of the mixed state with changes in the concentration, while $\mu_D$ arises due to change in {droplets'} surface energy and elastic energy upon change in its size.} 

The solution of equation \eqref{eq:quasi-eqm1} along with mass conservation constraint \eqref{eq:mass-conservation}, for each homogeneous region in the problem, fully describes the quasi-equilibrium state of the system, and serves as the initial condition for the long timescale dynamics of heterogeneous systems.\footnote{For a detailed treatment of the short timescale behavior from nucleation to growth of droplets, we refer the readers to \citep{kothari2020effect}.}

At the longer timescale, spatial heterogeneities in material properties can give rise to long-range migration in phase-separated systems.
{Rather than modeling the individual droplets, we consider a 1D aggregate representation with variations in droplet size and liquid concentration in the matrix represented by the two field variables $\phi_D(x)$ and $\phi(x)$, respectively. Such a 1D description is consistent with the symmetry of the problems we will study in this work.}
{Next, prescribing the flux $J$ {of liquid B} as proportional to its thermodynamic conjugate, i.e. gradients in chemical potential, we write}
\begin{equation}\label{eq:dynamics-flux}
  J = -\frac{D\phi}{kT} \bf{\nabla} \mu,
\end{equation}
where $D$ is the diffusivity of the liquid (upon linearization \eqref{eq:dynamics-flux} reduces to the commonly used Fick's law).

As the liquid starts to migrate across the matrix, the droplet-matrix equilibrium is disturbed $(\mu_D\neq\mu)$ and a cascading effect follows where droplet size may increase or decrease. 
This can be modeled by thinking of droplets as a source/sink term which gives the following equations for the dynamics of the system,
\begin{eqnarray}\label{eq:dynamics-pde}
    &\frac{\partial \phi(x,t)}{\partial t} + {\rm Div} J = s(x,t) \\
    &\frac{\partial \phi_D(x,t)}{\partial t} = -s(x,t).
\end{eqnarray}
Here, $s(x,t)$ is the source term that captures the behavior of the droplets, which can dissolve back into the matrix to replenish it or grow in size by absorbing the excess liquid in the matrix. This process is driven by the difference in chemical potential {of liquid B}; a thermodynamically consistent form of the source term is thus chosen as,
\begin{eqnarray}\label{eq:source}
     s(x,t) = \frac{K}{\nu_m}\left(\mu_D - \mu \right)H(\phi_D), \quad \mathcal{K} =\frac{ kT L^2K}{D\nu_m},
\end{eqnarray}
where $H(\phi_D)$ is the Heaviside function to ensure that the source is exhausted at $\phi_D = 0$, $L$ is a characteristic length scale of the system, and $K$ is a material property, which in its non-dimensionalized form, $\mathcal{K}$, is {called the \textit{dissolution number}} — it quantifies the relative eagerness of the droplet to give out the liquid to the matrix \citep{kothari2020effect}.
{The specific value of $K$ is not known and estimating it remains an area of future work.}
The kinetics of the process are governed primarily by the dissolution number --- a larger dissolution number makes the source term bigger and makes the droplets respond faster to any changes in the matrix. 

Equations \eqref{eq:dynamics-flux}-\eqref{eq:source}  complete the set of governing equations, which  must be supplemented with appropriate initial and boundary conditions to constitute a well-defined problem. 
In the following, we will apply the model to two cases of interest.

\section{Results and Discussion}\label{sec3}

\subsection{Case I - Size Regulation and Coarsening}\label{subsec2}
The size of various cellular organelles must be tightly regulated for their proper function. 
However, a distribution of droplets resulting from phase separation typically tends to coarsen over time, where larger droplets grow at the expense of smaller droplets --- a process known as Ostwald ripening. How do cells then achieve a stable distributions of multiple droplets? 
A few possible explanations have been proposed. 
For instance, chemical reactions taking place inside the droplets can potentially arrest the Ostwald ripening \citep{zwicker2015suppression}. 
Alternatively, certain cellular components can act as surfactants to stabilize the droplet size by reducing the surface energy of the droplet-matrix interface \citep{cuylen2016ki}. 
In the current work, we {explore} another mechanism that can stabilize the droplet distribution --- the elastic resistance to the growth and merger of droplets imposed by the medium.
{This mechanism was considered for phase separation in synthetic polymers in \citep{style2018liquid, kothari2020effect}. 
Here, we extend it to biopolymers and show that this mechanism can still be important at the scales relevant for the cells.}

Biopolymers often show strain-stiffening behavior, where they progressively stiffen as they are stretched.
Using our model and specializing $W(r)$ for strain-stiffening materials (see Appendix B), we study how elasticity regulates the size of droplets.
To illustrate the results, we select representative material values for chromatin -- a biopolymer found in nucleus -- from the literature \citep{ronceray2021liquid, zhang2020mechanical, berry2015rna}.
{Since the values of strain-stiffening parameter $n$ and the number density of droplets $n_d$ are not reported in these studies, we select $n$ to represent moderate strain-stiffening (with more values reported in the Appendix D), and choose $n_d$ to be proportional to the elastic modulus, as motivated by the experimental finding from Style et al. \citep{style2018liquid}. Accordingly, we set  $n_d = \alpha E$, where the choice of $\alpha = 10^{13}$N$^{-1}$m$^{-1}$ is made to obtain droplet sizes in ballpark of those reported in \citep{zhang2020mechanical}. }

\begin{figure}[ht!]
\centering
  \makebox[0.45\textwidth][c]{\includegraphics[width=0.4\textwidth]{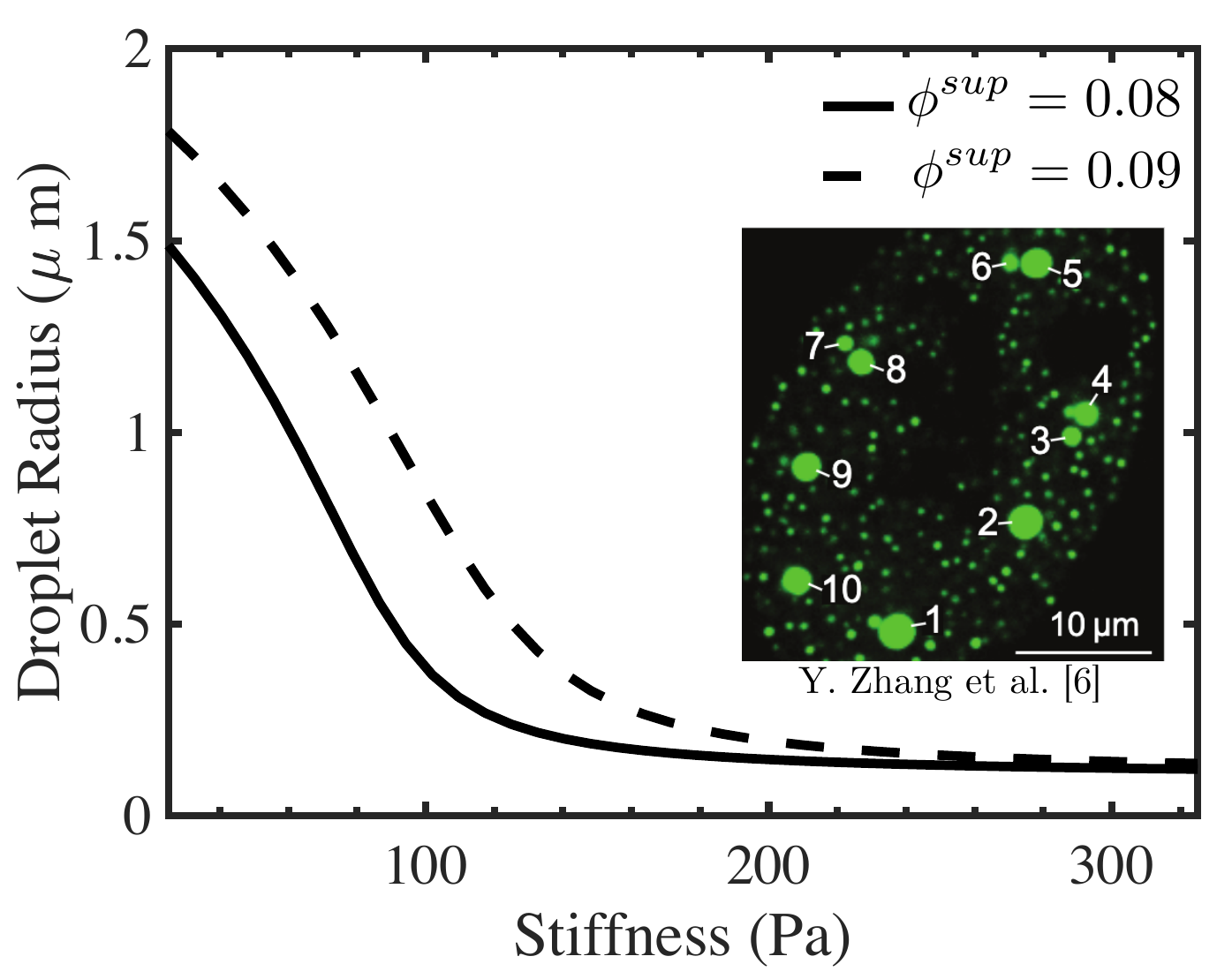}} 
  \caption{{\textbf{Typical droplet size as a function of chromatin stiffness at short timescales.} Material Properties: $n=0.95, \Gamma = 5\times10^{-7}$ Nm$^{-1}$, $\nu_m = 1\times10^{-23}$ m$^{3}$, $\alpha = 10^{13}$ N$^{-1}$m$^{-1}$, $\chi = 2.14$. Inset: optogenetically nucleated droplets in chromatin network show typical size of a few microns \citep{zhang2020mechanical}.}} \label{fig:droplet_size}
\end{figure}

The droplet sizes at short timescales (or in absence of heterogeneity that can lead to long timescale migration) are shown in Fig. \ref{fig:droplet_size} for two different supersaturations and moderate strain-stiffening.
In contrast to the commonly assumed neo-Hookean response, in a strain-stiffening matrix the elastic energy cost of expanding  a single large droplet is higher than growing multiple small droplets for the same total volume. Thus, strain-stiffening provides a clear explanation for the observed multi-droplet state.

\begin{figure}[h!]%
\centering
\includegraphics[width=0.48\textwidth]{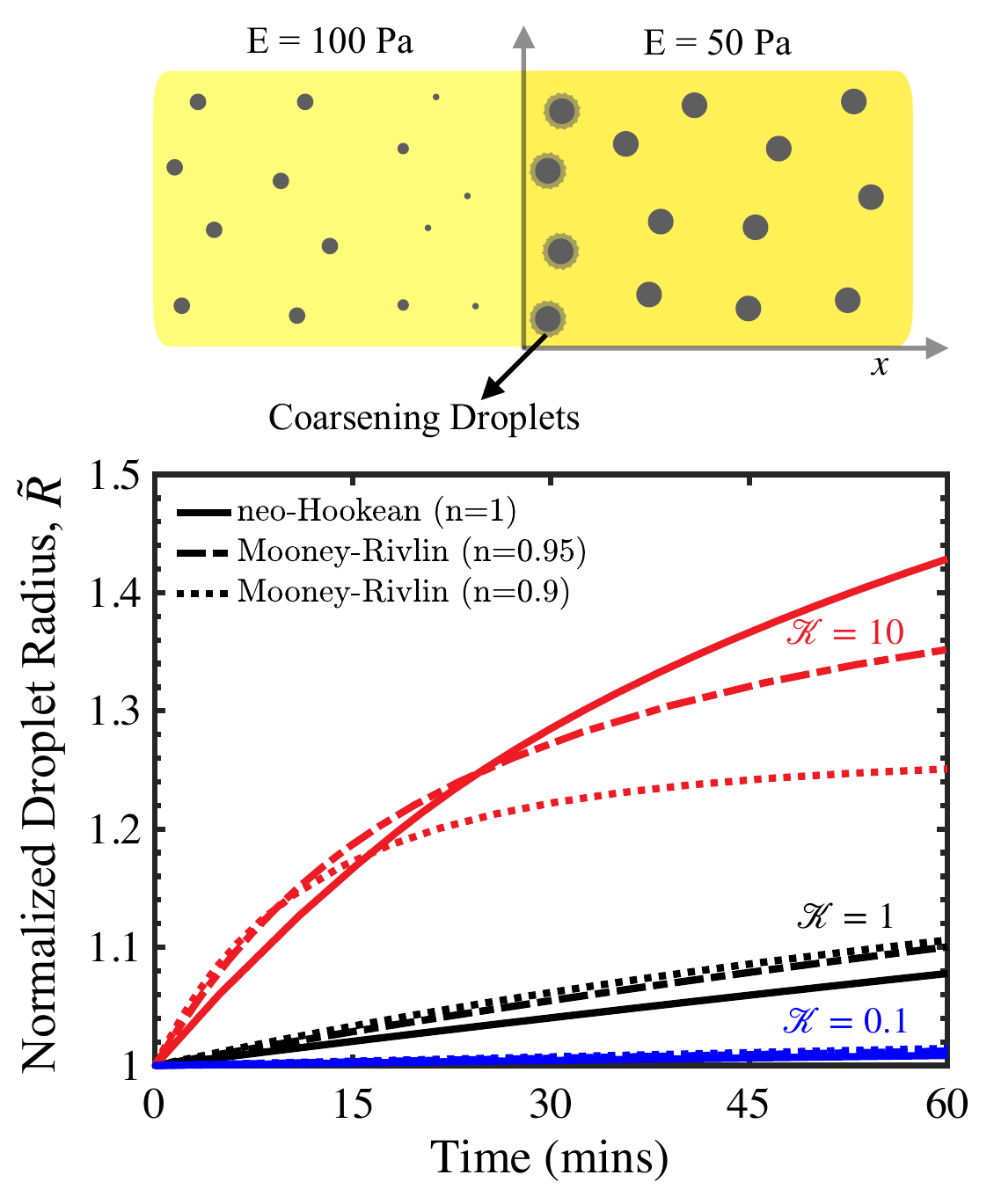}
\caption{{\textbf{Elasticity-driven coarsening in heterogeneous systems.} Normalized droplet radius, $\tilde{R}$, for droplets at the interface on the right side. Material Properties: $E = 100$ Pa (Left); $E = 50$ Pa (Right); $\Gamma = 5\times10^{-7}$ Nm$^{-1}$, $\nu_m = 5\times10^{-24}$ m$^{3}$, $D = 5\times10^{-11}$m$^{2}$s$^{-1}$, $\alpha = 10^{13}$ N$^{-1}$m$^{-1}$, $\chi = 2.14$ (both sides).}}\label{fig:coarsening}
\end{figure}

\begin{figure*}[h!]%
\centering
\includegraphics[width=\textwidth]{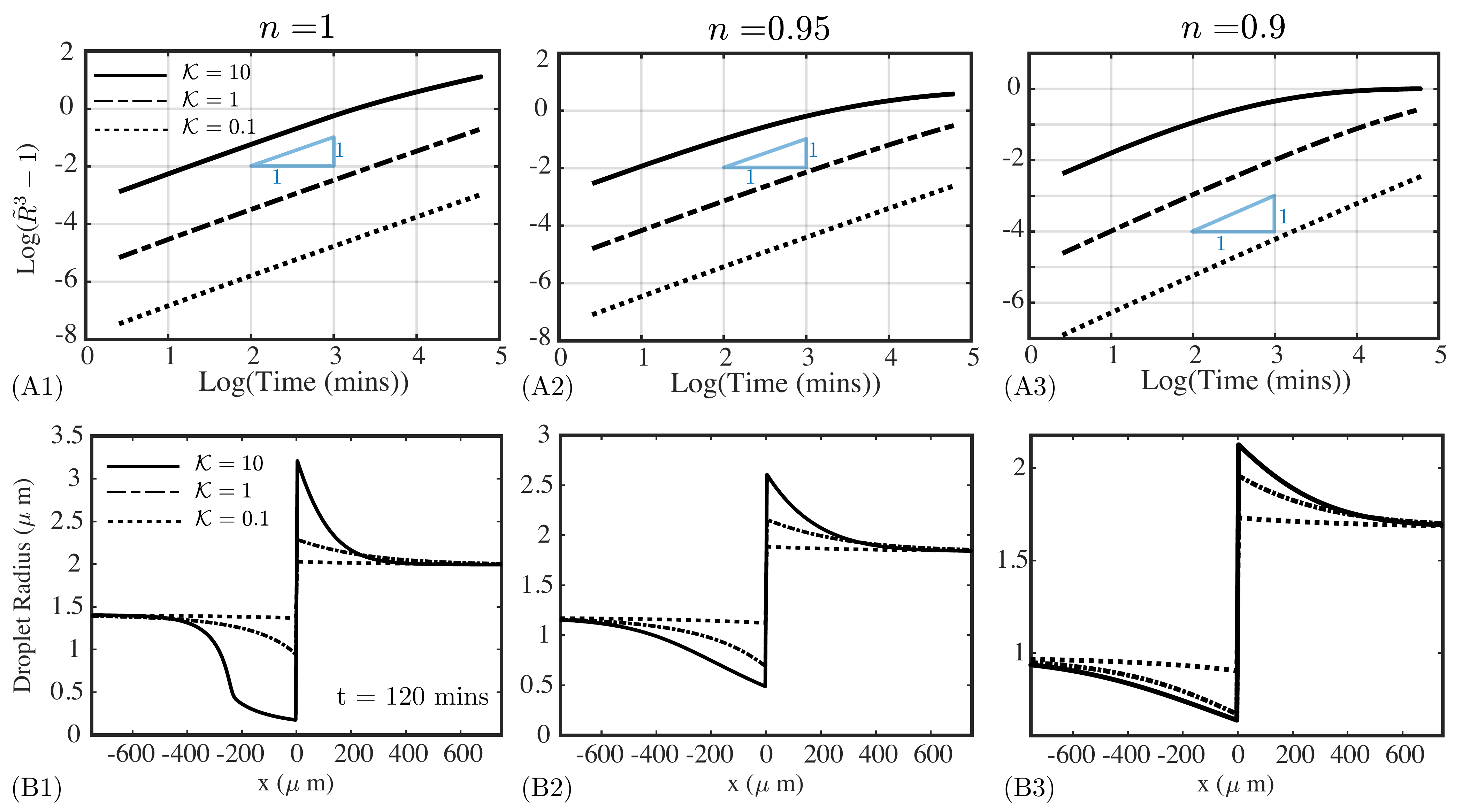}
\caption{{\textbf{Dynamics of coarsening in heterogeneous systems.} Same system as shown in the schematic in Fig. 2. Material Properties: $E = 100$ Pa (Left); $E = 50$ Pa (Right); $\Gamma = 5\times10^{-7}$ Nm$^{-1}$, $\nu_m = 5\times10^{-24}$ m$^{3}$, $D = 5\times10^{-11}$m$^{2}$s$^{-1}$, $\alpha = 10^{13}$ N$^{-1}$m$^{-1}$, $\chi = 2.14$ (both sides). (A) Rate of coarsening for different values of strain stiffening and dissolution numbers; $\tilde{R}$ denotes the normalized droplet radius. Rate of Ostwald ripening is shown for comparison in blue. (B) Spatial distribution of droplet sizes for $t= 120$ mins. Also see Appendix E for coarsening behavior at longer times.}}\label{fig:coarseningdynamics}
\end{figure*}

We find that even before onset of long timescale effects, the response of the system is highly nonlinear in its dependence on the constitutive properties. 
{By choosing a reasonable value for $\alpha$, the droplet sizes obtained from the model are in good agreement with the experimentally observed range of droplet sizes \cite{zhang2020mechanical}. This   supports the hypothesis that elastic resistance from the medium can play an important role in size regulation of the droplets. However, experimental measurements of   $n$ and $n_d$ are required to provide accurate predictions}.
Beyond the intuitive trends that are captured (i.e. that a stiffer medium results in smaller droplets), the rapid decay of droplet radius, from a maximum value to a nearly unaffected size, indicates a narrow zone of high sensitivity within a range of stiffnessses that are accessible to active biological materials. 
Namely, biological systems can use this process as a \textit{mechanical switch}, whereby activating relatively small changes in stiffness can induce large changes in droplet size. (Also see Appendix D for strain-stiffening effects.)

While at the long timescale the surface tension driven coarsening is hindered by the presence of elastic networks, these systems may show limited coarsening driven by gradients in stiffness.
To study this coarsening behavior in an elastically heterogeneous medium, {we take inspiration from \citep{rosowski2020elastic} and} construct the simplest heterogeneous setup by considering two homogeneous connected regions of differing stiffnesses, as shown in Fig. \ref{fig:coarsening}.
{We require continuity of flux at the interface, and consider and infinite domain (i.e. $x\in(-\infty,\infty)$).}
After the initial nucleation and growth of droplets is complete, a longer timescale transport emerges that causes the liquid to flow along the direction of decreasing chemical potential.

Fig. \ref{fig:coarsening} shows the results for the normalized droplet size {at the interface, denoted by $\tilde{R}$,} evolving with time, {where we use the initial droplet radius -- at the end of the short timescale -- to normalize: $\tilde{R} = r(x=0^+, t)/r(x=0^+, t=0)$.}
For strain-stiffening media, the droplets  eventually reach a final size
{(Figs. E3 and E4 in Appendix E show coarsening plots up to 1200 minutes that show the size convergence).}
{Figs. \ref{fig:coarseningdynamics}(A1-A3) show the rate of coarsening for a range of strain-stiffening parameters and dissolution numbers, {where the blue slopes show the Ostwald ripening rate for comparison.}
At the initial times, the coarsening rate follows the Ostwald ripening power law ($\Tilde{R}^3 -1 \propto t$), and diverges from it at longer times. 
Cases with $\mathcal{K}=10$ are the first to show a significant deviation because of their faster dynamics, whereas the cases with smaller $\mathcal{K}$ will take longer to diverge.
In particular, we find that even for moderate levels of strain-stiffening (i.e. Figs. \ref{fig:coarseningdynamics}(A2, A3)), significant deviation from Ostwald ripening can occur and the coarsening is eventually arrested.
For a neo-Hookean material $(n=1)$, the coarsening continues indefinitely. Nonetheless,  it  eventually becomes slower than Ostwald ripening.}
{Finally, Figs. \ref{fig:coarseningdynamics}(B1-B3) show the spatial distribution of droplet sizes for all the different cases at $t =120$ minutes.}
{Fig. \ref{fig:coarsening} and Figs. E2-E4 (in Appendix E) also highlight the role of the dissolution number: since it controls the dynamics of matrix-droplet transport, a higher dissolution number leads to faster coarsening, and the droplets arrive at their final size faster (in the case of a strain-stiffening medium).}

\subsection{Case II - Spatial Localization of Droplets in Finite Domains}\label{subsubsec2}

Cells can spatially localize condensates by employing phase separation. 
{For instance, Brangwynne et al. \citep{brangwynne2009germline} showed that in \textit{Caenorhabditis elegans}, P granules, starting from a uniformly distributed state, localize to the posterior of the $\sim 50 \mu m$-long cell over a period of $\sim$ 10 minutes.}
Cells are also highly dynamic and regulate their internal structure as well as mechanical properties both spatially and temporally \citep{heidemann2004towards}.
These two observations naturally raise the question: how does the change of stiffness in the {cellular} medium impact the localization of the droplets\footnote{{While phase separation in \citep{brangwynne2009germline} was shown to be controlled by concentration of polarity proteins, we use spatiotemporal localization of condensates as a motivating example to explore the role of elasticity.}}?
To answer this question, we consider a scenario where the system develops a stiffness gradient over time and study its impact on the localization of droplets.
Initially, the system is homogeneous and has a uniform distribution of droplets, the size of which can be obtained by solving equations \eqref{eq:mass-conservation} and \eqref{eq:quasi-eqm1}.
Over a timescale of minutes, the system develops a linear gradient in stiffness, where the stiffness of the left end {($x = -L/2$)} starts to increase, while that of the right end {($x = L/2$)} remains fixed: {$E(x,t) = 100 - 10(x/L-1/2)t$ Pa, where $t$ is the time in minutes and $x \in [-L/2,L/2].$}
By establishing a stiffness gradient, the system alters the chemical potential of the droplets, which now exceeds that of the matrix, thus disturbing the equilibrium and causing the liquid to migrate from left to right (along the decreasing stiffness). 
{We apply a \textit{no flux} boundary condition on both the ends (i.e. $J=0$ at $x=-L/2, L/2$).
{Note that we take $n_d$ to be constant during this process, corresponding to $E = 100$Pa.}
In the initial process of nucleation and growth, the regions near the droplet-matrix interface experience high strains and, thus, inevitable damage. These locations also serve as preferential spots for droplet growth. Furthermore, the elastic networks prevent the droplets from coalescing. Thus, once the short timescale equilibrium has been achieved, $n_d$ remains the same.}
\begin{figure}[ht]
\centering
  \makebox[0.5\textwidth][c]{\includegraphics[width=0.48\textwidth]{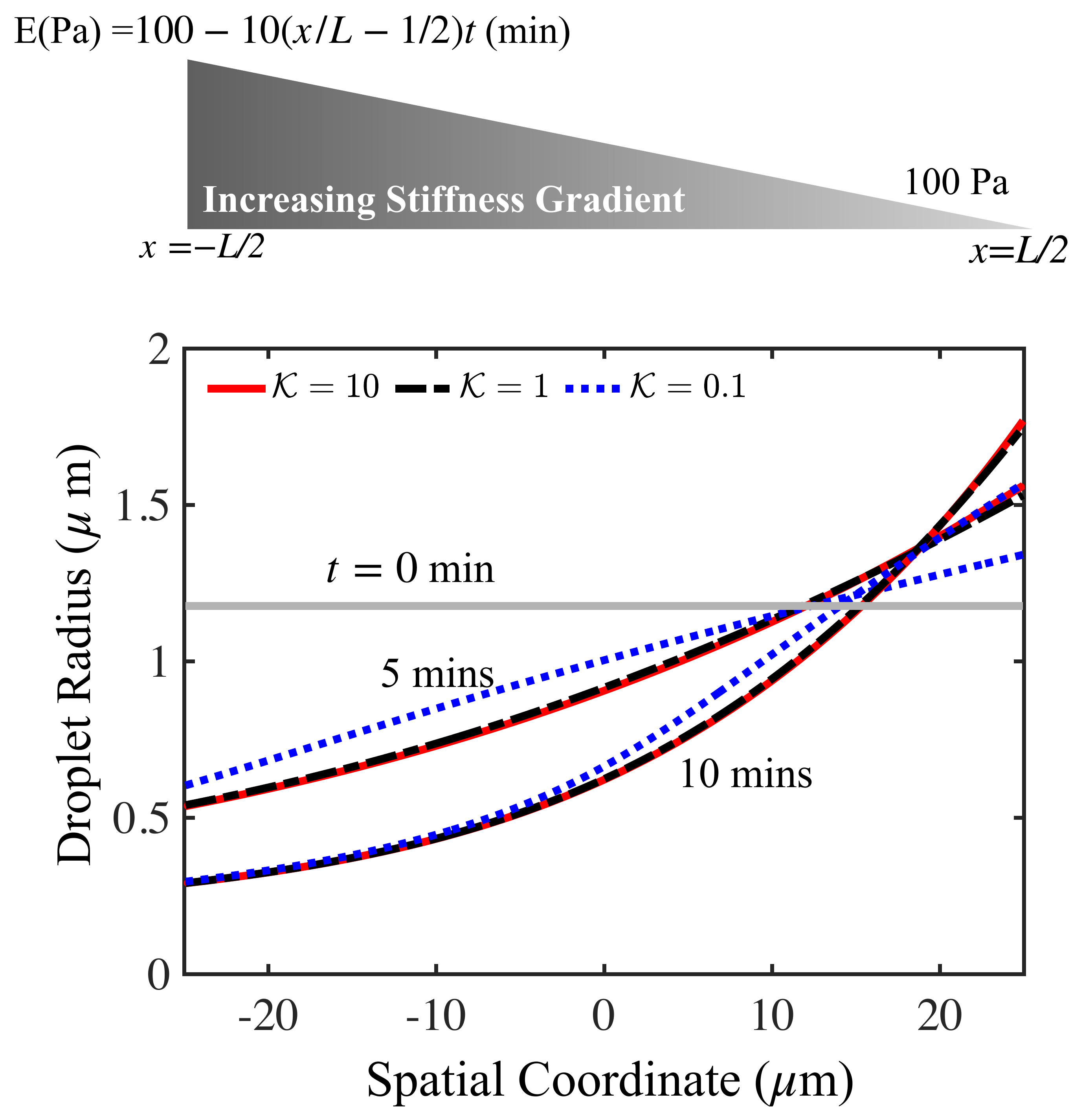}}
 { \caption{\textbf{Spatial localization of droplets by stiffness gradient.} A linear stiffness gradient is established across a 50 $\mu$m region, which is previously homogeneously stiff. Material Properties: $E(x,t) = 100 - 10(x/L-1/2)t$ Pa where $t$ is in minutes, $n_d = 10^{15}$ m$^{-3}$, $\Gamma = 5\times10^{-7}$ Nm$^{-1}$, $\nu_m = 5\times10^{-24}$ m$^{3}$, $D = 5\times10^{-11}$m$^{2}$s$^{-1}$, $\chi = 2.14$, $n = 0.95$.} }\label{fig:stiffness_gradient}
  
\end{figure}

Fig. 4 shows the results of this process over time for {a $50 \mu m$-long domain}. We find that the droplets localize over biologically relevant timescales ($\sim$ 10 mins) to the {right side.} 
We also see that a higher dissolution number promotes faster localization. 
{When $\mathcal{K}$ is sufficiently large, the droplet-matrix system can quickly adapt to changes in the matrix stiffness and continuously maintain a quasi chemical equilibrium. If $\mathcal{K}$ is too low, the droplet dynamics will lag behind the stiffness changes in the matrix, and the system will always be out of chemical equilibrium.}
This effect can be especially important in the tightly regulated cellular environments. 
We speculate that the dissolution number lumps together factors like local stress, swelling of the network, and diffused damage around the droplet \citep{kim2020extreme}; a multi-scale approach can better resolve these local dynamics and is beyond the scope of the current work.

\section{Conclusion}\label{sec13}

In conclusion, these quantitative predictions in a simple and experimentally realizable setting are aimed at uncovering the role of elastic driving forces in intracellular phase separation.
Even at the cellular length scales, the elastic resistance from the cellular media can impact the growth and dynamics of biomolecular condensates -- both of which are critical to the functioning of cells.
While we have established that elasticity-driven mechanisms are accessible to biological systems, the question of whether and to what extent cells use these mechanisms remains an open question.
Future work in this area must focus on understanding if the changes in cellular stiffness are fast and significant enough to affect the outcome of LLPS.
The recent development of advanced optogenetic techniques to selectively initiate phase separation \cite{Spatio2017}, together with the capability to create spatially heterogeneous crosslinking can be a promising technique to quantify the dynamics arising due to differences in elastic properties, and to test the predictions of our model. {Finally, the model presented here is not without limitations. Future work should extend beyond 1D geometries to reveal more complex phase separation phenomena, and should account for additional physical mechanisms that may become important in cellular systems, such as strain dependent surface energy  \citep{jensen2017strain, xu2017direct}.}

\backmatter

\bmhead{Acknowledgments}
The authors acknowledge the support of the Office of Naval Research, United States of America and Dr. Timothy B. Bentley, Program Manager, under award number N00014-20-1-2561.

\section*{Declarations}

The authors declare no conflict of interests.

 \begin{appendices}

 \section{Flory-Huggins Parameter}\label{secA1}

{The Flory-Huggins parameter $\chi$ can be derived  for a given solubility of liquid B in the mixture, $\phi^{sat}$, as \cite{kothari2020effect}, 
\begin{equation}
      \chi = -\frac{\log\phi^{sat} + (1-\phi^{sat})}{(1-\phi^{sat})^2},
\end{equation}
 in the absence of any elastic and surface effects. Throughout this paper, we use $\chi = 2.14$ which translates to a solubility concentration of $\phi^{sat} =0.06$.}

\section{Strain Energy of Droplet Growing Inside a Strain-Stiffening Elastic Matrix}\label{S1}

To capture nonlinear material response at large strains, we use the incompressible Mooney-Rivlin constitutive model for the elastic response of the matrix \citep{kothari2020effect}. The work done in expanding a single droplet from stress-free radius $r_0$ to an expanded radius $r$ is given as,
\begin{eqnarray} \label{eq:elastic-work}
\begin{split} 
  W(r) = nE\bigg(\frac{5}{6}-\frac{r_0}{r}- \frac{r_0^3}{3r^3 }+ \frac{r_0^4}{2r^4} \bigg)+ \\ (1-n)E\bigg(\frac{r}{2r_0}-\frac{1}{3} -\frac{r_0^2}{r^2}  + \frac{5}{6}\frac{r_0^3}{r^3}\bigg),
\end{split}
\end{eqnarray}
where $E$ is the stiffness of the crosslinked polymer, and $0 \leq n\leq 1$ is the strain-stiffening parameter; $n=1$ represents the neo-hookean material (no strain stiffening) and the level of strain stiffening increases with decreasing $n$. 
{The stress-free radius is the length scale at which the elastic strain energy dominates over the surface energy in the growth of the droplet, and throughout this paper we choose $r_0=0.1 \mu m$.}
As explained in \citep{kothari2020effect}, $r_0$ may be different from the pore size of the elastic network in the matrix.
{We also note that our choice of $r_0$ is consistent with [6]: Fig. 3(b) in [6] shows that in a cross-linked matrix, the pressure initially drops as the droplet grows, before increasing again when the radius is $\sim 0.1 \mu m$, which identifies the region of elastic resistance from chromatin to the growth of droplets.}

\section{Minimization of Free Energy}\label{S0}
Following \eqref{eq:total-free-energy}, the total free energy can be written as,
\begin{eqnarray}
\begin{split}
   \Delta \bar{G}(\phi,\phi_D) =  (1-\phi_D)\frac{k T}{\nu_m}\{\phi \ln\phi+ \chi(T) \phi(1-\phi)\} \\+ 4\pi r^2\Gamma {n_d} + \frac{4\pi}{3}r^3W(r) {n_d}.
 \end{split}
 \end{eqnarray}
 
The system also satisfies the mass conservation constraint,
\begin{equation}
  \phi^{sup} = \phi(1-\phi_D) +\phi_D.
\end{equation}
The energy minimization condition
\begin{equation}
    \frac{{\rm d}\Delta \bar{G}}{{\rm d}\phi} = 0
\end{equation}
can then be evaluated and simplified using the following relations: $\frac{\partial\left(1-\phi_D\right)}{\partial \phi} = \frac{(1-\phi_{D})}{(1-\phi)}$ and $\phi_D = \frac{4\pi}{3}r^3 n_d$,

\begin{eqnarray}
\begin{split}
  &\frac{{\rm d}\Delta \bar{G}_{mix}}{{\rm d}\phi} = \frac{k T}{\nu_m}\frac{\partial\left(1-\phi_D\right)}{\partial \phi}\{\phi \ln\phi+ \chi(T) \phi(1-\phi) \} \\ & +\frac{k T}{\nu_m}(1-\phi_D)\{1+ \ln\phi+ \chi(T)(1-2\phi)\}\\
    =&\frac{k T}{\nu_m}\frac{\left(1-\phi_D\right)}{(1-\phi)}\{ \ln\phi+ (1-\phi) + \chi(T) (1-\phi)^2 \} \\ =& \frac{\left(1-\phi_D\right)}{(1-\phi)}\frac{\mu}{\nu_m}
\end{split}
 \end{eqnarray}

\begin{eqnarray}
\begin{split}
    &\left(\frac{{\rm d}\Delta \bar{G}_{sur}}{{\rm d}\phi} + \frac{{\rm d}\Delta \bar{G}_{el}}{{\rm d}\phi}\right) = \\& \frac{\partial}{\partial \phi_D}\left\{\phi_D\left( \frac{3\Gamma}{r}+ W(r) \right) \right\}\frac{\partial \phi_D}{\partial\phi}\\
    =& - \frac{\left(1-\phi_D\right)}{(1-\phi)}\left\{\frac{2\Gamma}{r} + W(r) +\frac{r}{3}W'(r) \right\} \\
    =& - \frac{\left(1-\phi_D\right)}{(1-\phi)}\frac{\mu_D}{\nu_m}
\end{split}
 \end{eqnarray}

Finally, these equations yield 
\begin{equation}
    \frac{{\rm d}\Delta \bar{G}}{{\rm d}\phi} = \frac{\left(1-\phi_D\right)}{(1-\phi)}\left(\frac{\mu -\mu_D}{\nu_m}\right) =0
\end{equation}
as the equilibrium condition shown in \eqref{eq:quasi-eqm1}.

\section{Equilibrium Droplet Size Variation with Strain-Stiffening} \label{S2}
\begin{figure}[h]
  \includegraphics[width=0.5\textwidth]{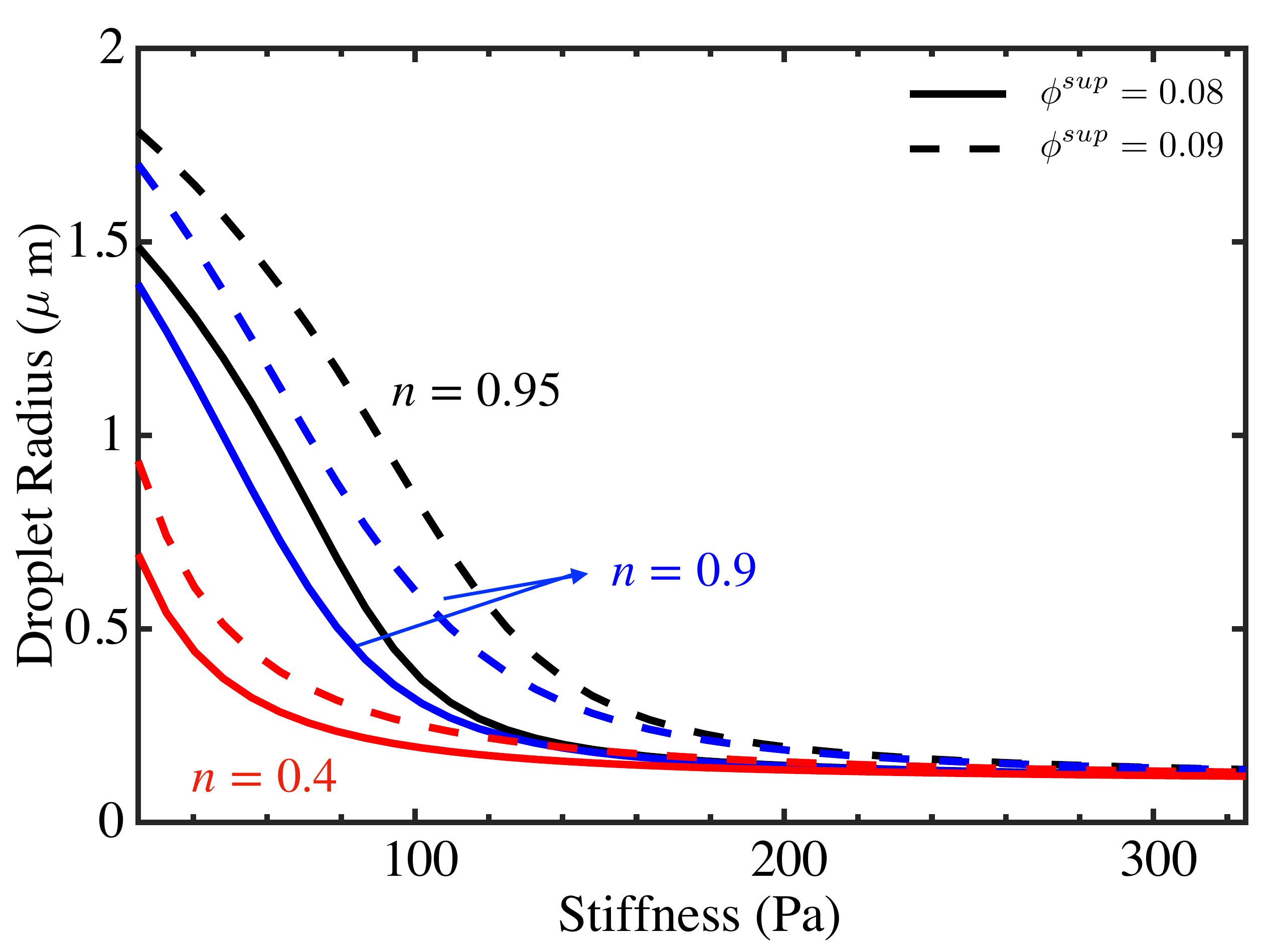}
 \caption{Equilibrium droplet sizes for different levels of strain stiffening.}
 \label{fig:SI_droplet_sizes}
\end{figure}

The equilibrium droplet size is determined by the solution of \eqref{eq:quasi-eqm1} together with the mass conservation constraint \eqref{eq:mass-conservation}.
Using the form outlined in \eqref{eq:elastic-work} for $W(r)$, we study the sensitivity of the equilibrium droplet sizes to the level of strain stiffening, as governed by the parameter $n$. 
Figure \ref{fig:SI_droplet_sizes} shows that the increasing strain-stiffening decreases the equilibrium droplet size. 

 \section{Coarsening Plots for Longer Times} \label{S3}
{The following figures show normalized droplet radius, $\tilde{R}$, over time, for droplets at the interface on the right side. The three figures are for $n = 1, 0.95$ and $0.9$ respectively. Material Properties: $E = 100$ Pa (Left); $E = 50$ Pa (Right); $\Gamma = 5\times10^{-7}$ Nm$^{-1}$, $\nu_m = 5\times10^{-24}$ m$^{3}$, $D = 5\times10^{-11}$ m$^{2}$s$^{-1}$, $\alpha = 10^{13}$ N$^{-1}$m$^{-1}$, $\chi = 2.14$ (both sides) which translates to a solubility concentration of $\phi^{sat} = 0.06$ in the absence of elastic and surface effects. In strain-stiffening media, the coarsening is arrested, leading to final size as is evident from Figs. E3 and E4. The systems with smaller $\mathcal{K}$ take longer time to reach the final state. }
\begin{figure}[h]
 \includegraphics[width=0.5\textwidth]{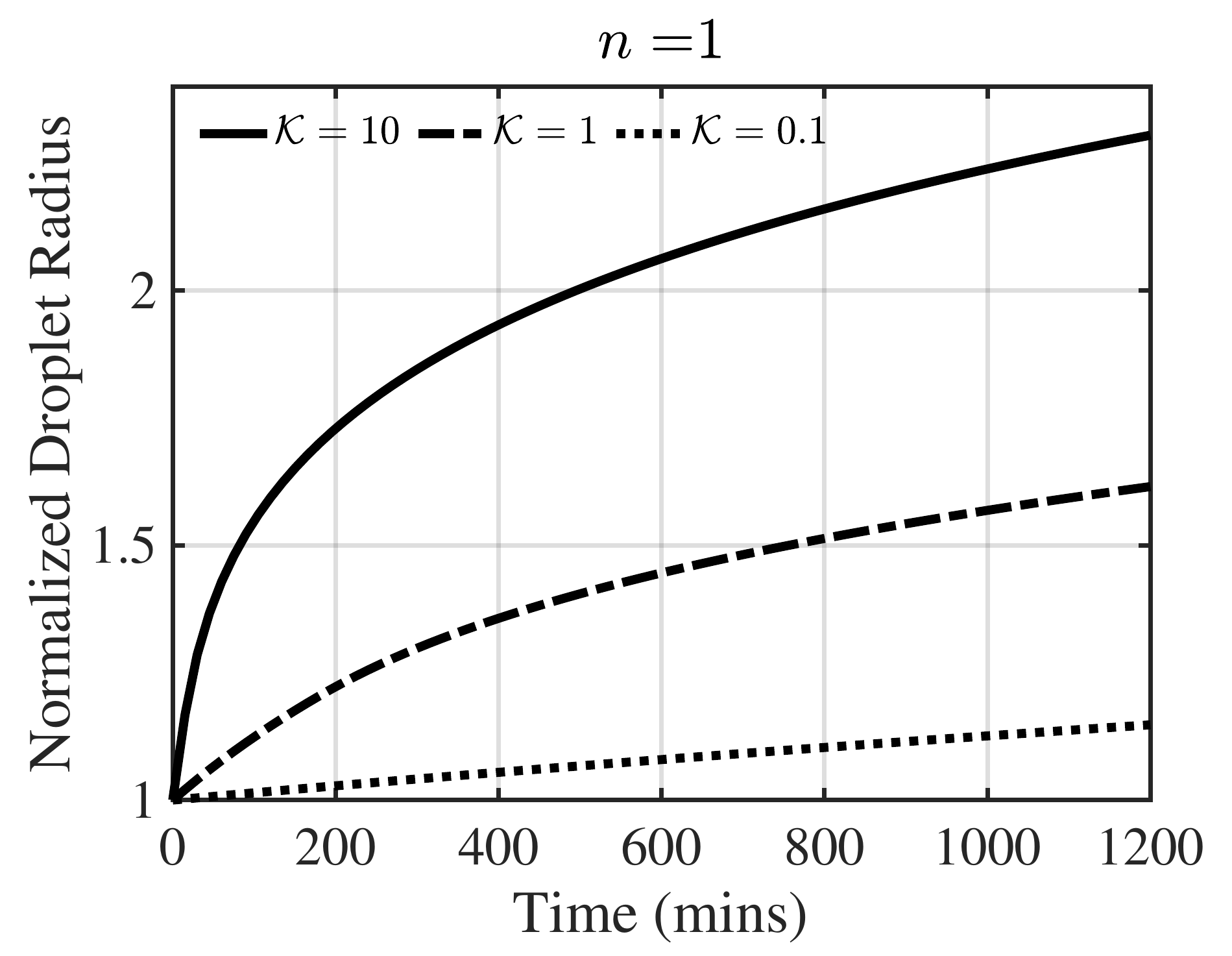}
 \caption{{Coarsening dynamics for $n = 1$.}}
 \label{fig:coarsening1}
\end{figure}

\begin{figure}
\includegraphics[width=0.5\textwidth]{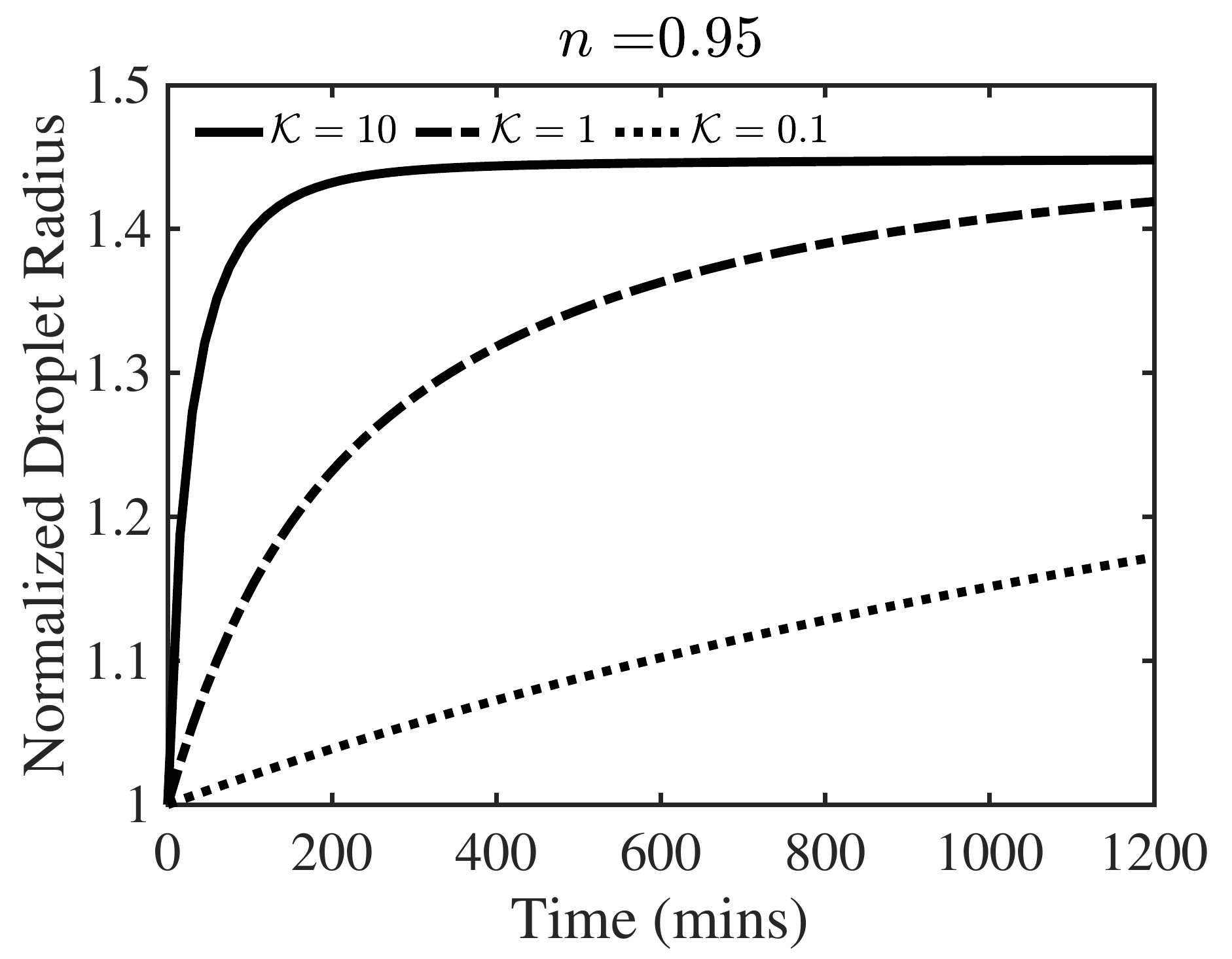}
  \caption{Coarsening dynamics for $n = 0.95$.}
  \label{fig:coarsening2}
\end{figure}

\begin{figure}
  \includegraphics[width=0.5\textwidth]{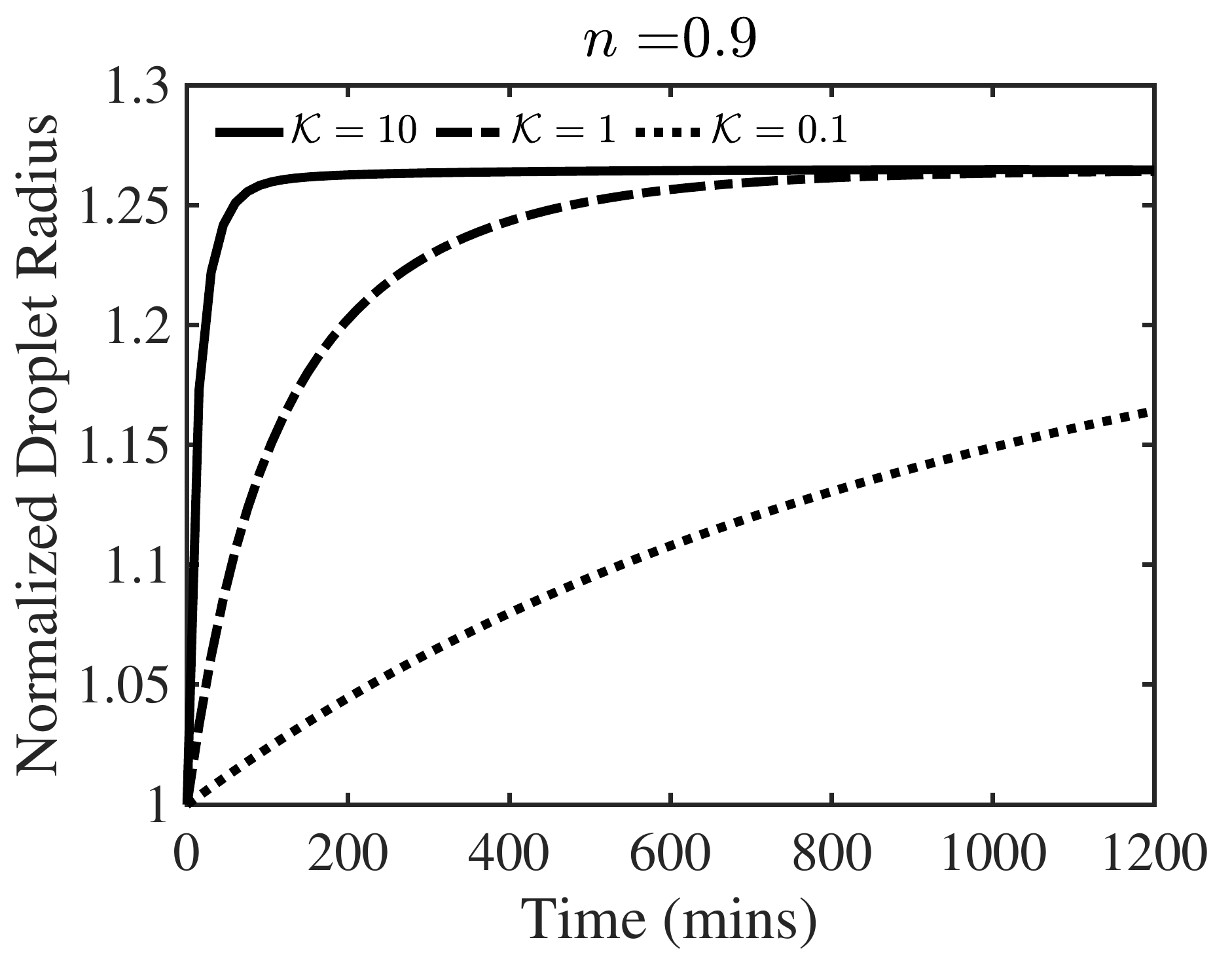}
  \caption{Coarsening dynamics for $n = 0.9$.}
  \label{fig:coarsening3}
\end{figure}
\hfill

\clearpage

\clearpage
\end{appendices}
\bibliography{myfile}


\end{document}